\documentclass[debug]{rmaa}

\usepackage{paralist}
\usepackage{natbib}
\usepackage{psfrag,color}
\usepackage{multirow}

\usepackage{lipsum}
\usepackage{url}
\usepackage{aas_macros}
\usepackage{array}
\usepackage{amsmath}
\usepackage{booktabs}
\usepackage{amsthm}
\usepackage{caption}
\usepackage{tabularx} 
\usepackage{setspace}
\usepackage{hyperref} 
\usepackage{longtable}
\usepackage{geometry}
\usepackage{graphicx}
\usepackage{epstopdf}
\usepackage {float}
\usepackage{supertabular,booktabs}

\newcommand{\kms}{km\,s$^{-1}$}

\SetYear{2025}

\title{Spectroscopic study of blue straggler stars in the Globular Cluster NGC 3201}

\author{
  Gourav Kumawat\altaffilmark{1,2}, Arvind K. Dattatrey\altaffilmark{3}, R.K.S. Yadav\altaffilmark{3}}

\altaffiltext{1}{Indian Institute of Science Education and Research, Bhopal, MP, INDIA.}
\altaffiltext{2}{University of Alberta, Edmonton, AB, CANADA}
\altaffiltext{3}{Aryabhatta Research Institute of Observational Sciences, Nainital, UK, INDIA}

\suppressfulladdresses

\listofauthors{Gourav Kumawat, Arvind K. Dattatrey, R.K.S. Yadav}
\indexauthor{Kumawat, G}
\indexauthor{Dattatrey, A}
\indexauthor{Yadav, R.K.S.}

\addkeyword{Blue Straggler Stars}
\addkeyword{Globular star clusters: NGC 3201}
\addkeyword{Spectroscopy }
 
\shortauthor{G. Kumawat et al.}
\shorttitle{BSS in NGC 3201}

\abstract{We conducted a spectroscopic study of 39 blue straggler stars in the globular cluster NGC 3201. The spectra of these stars were collected from the literature. We determined the radial velocity, atmospheric parameters (T$_{eff}$, log $g$),  and the abundance of Mg, as well as the metallicity ([Fe/H]) of the blue straggler population. The mean radial velocity and [Fe/H] are determined to be 498.0$\pm$5.3 km/s and -1.42$\pm$0.27, respectively, for the blue straggler stars. The derived [Fe/H] is consistent, within uncertainties, with the cluster's [Fe/H] of –1.59 dex. The mean [Mg/Fe] for the blue straggler stars is estimated to be 0.36$\pm$0.73. Importantly, this study first estimates [Mg/Fe] for the blue straggler stars in the cluster NGC 3201.}

\begin{document}
\maketitle

\section{Introduction} \label{introduction}

Globular clusters (GCs) are among the oldest stellar systems, consisting of thousands of stars located at the same distance from us and moving together in space. Due to their high stellar densities, frequent gravitational interactions occur within them, leading to several dynamical processes that give rise to exotic stellar populations. One such population is that of Blue Straggler stars (BSSs).

\begin{table*}
\centering 
\caption{ID\lowercase{s}, RA, DEC, Visual Magnitude, and Membership Probability of the 39 BSS\lowercase{s} Used in This Study (Refer to Text for Details)}


\label{tab:bss_crossmatch_results}
\begin{tabular}{cccccc}
\toprule 
BSS ID & Cosmic-Lab ID & RA (degrees)          & DEC (degrees)         & m$_V$   & PROB (\%)\\
\midrule
B1	&	1023539	&	154.3993515	&	-46.4078560	&	17.54	&	97	\\
B2	&	1013700	&	154.3962520	&	-46.4182565	&	17.27	&	96.9	\\
B3	&	1025211	&	154.3849299	&	-46.4076830	&	17.58	&	97.3	\\
B4	&	1021152	&	154.4195717	&	-46.4093147	&	17.17	&	97.3	\\
B5	&	1025619	&	154.3816172	&	-46.4041477	&	16.86	&	99.9	\\
B6	&	1031149	&	154.4004033	&	-46.3970078	&	16.69	&	97.7	\\
B7	&	1010831	&	154.4189046	&	-46.4216023	&	16.34	&	97.3	\\
B8	&	1021331	&	154.4184729	&	-46.4027317	&	17.39	&	97.7	\\
B9	&	1009786	&	154.4276377	&	-46.4137044	&	16.95	&	97.7	\\
B10	&	1020369	&	154.4272846	&	-46.4087656	&	17.20	&	97.1	\\
B11	&	1031222	&	154.3994121	&	-46.3927124	&	17.58	&	97.7	\\
B12	&	1026151	&	154.3766105	&	-46.4000632	&	17.77	&	97.2	\\
B13	&	1016643	&	154.3700469	&	-46.4114471	&	17.22	&	97.1	\\
B14	&	1006762	&	154.3834415	&	-46.4321004	&	16.60	&	97.3	\\
B15	&	1009269	&	154.4330228	&	-46.4137157	&	16.64	&	97.5	\\
B16	&	1016923	&	154.3668729	&	-46.4150062	&	17.82	&	97.4	\\
B17	&	1002749	&	154.4246075	&	-46.4282995	&	17.60	&	97.3	\\
B18	&	1008650	&	154.4390510	&	-46.4182480	&	17.45	&	95.3	\\
B19	&	1030901	&	154.4032020	&	-46.3843063	&	17.47	&	100	\\
B20	&	2716018	&	154.3760880	&	-46.4361699	&	17.19	&	100	\\
B21	&	2711709	&	154.4041397	&	-46.4461522	&	17.74	&	0.06	\\
B22	&	1035238	&	154.3842813	&	-46.3772396	&	17.38	&	100	\\
B23	&	2714632	&	154.3860264	&	-46.4501061	&	16.43	&	100	\\
B24	&	2705506	&	154.4472108	&	-46.3887971	&	17.54	&	99.9	\\
B25	&	2714006	&	154.3898416	&	-46.3689811	&	17.51	&	100	\\
B26	&	2709841	&	154.4152943	&	-46.3652141	&	17.59	&	99.9	\\
B27	&	2720497	&	154.3341108	&	-46.4439012	&	16.49	&	100	\\
B28	&	2710692	&	154.4103191	&	-46.4691207	&	17.36	&	100	\\
B29	&	2711836	&	154.4034550	&	-46.4747336	&	17.25	&	99.9	\\
B30	&	2600533	&	154.3140993	&	-46.3867423	&	17.19	&	100	\\
B31	&	2702518	&	154.4824695	&	-46.3775296	&	16.63	&	100	\\
B32	&	2707141	&	154.4338850	&	-46.3495438	&	17.25	&	96.5	\\
B33	&	2717272	&	154.3667567	&	-46.4837663	&	17.85	&	99.9	\\
B34	&	2601851	&	154.2893054	&	-46.4032201	&	17.46	&	100	\\
B35	&	2708408	&	154.4249930	&	-46.3192117	&	17.42	&	100	\\
B36	&	2603838	&	154.2324762	&	-46.3496011	&	17.36	&	100	\\
B37	&	2714749	&	154.3855481	&	-46.5545236	&	17.20	&	100	\\
B38	&	2301092	&	154.2432987	&	-46.2935999	&	16.85	&	0	\\
B39	&	2801544	&	154.6364229	&	-46.4157168	&	17.64	&	100	\\
\bottomrule
\end{tabular}
\end{table*}

\begin{figure*}
    \centering
    \includegraphics[scale = 0.09]{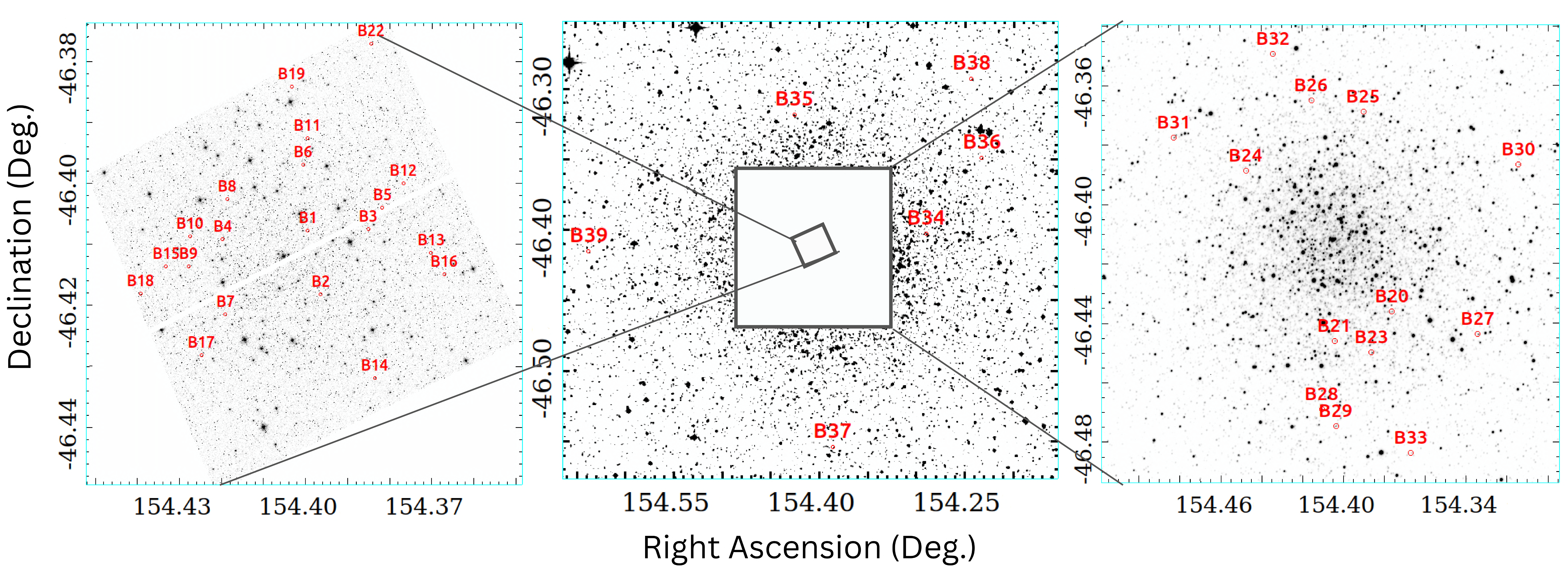}
    \caption[NGC 3201 BSSs locatios]{Spatial locations of the BSSs in different regions of NGC 3201. The BSSs are marked with red circles and labelled with their IDs in all the images.  
    \textbf{Left:} HST F606W image of the core of NGC 3201 taken by the HST ACS Wide Field Camera on March 14, 2006, with a total exposure time of 5 seconds \citep{Nadiello2018}. \textbf{Middle:} UK Schmidt telescope image of NGC 3201 taken on March 4, 1978, with a total exposure of 4200 seconds sourced from the DSS survey produced at the Space Telescope Science Institute (STScI).\textbf{Right: } ESO/MPG 2.2m telescope image of NGC 3201 in the V/89 filter taken by wide-field imager (WFI) with a total exposure of 480 seconds (credits: ESO Imaging Survey). }
    \label{fig:ngc3201_bss_loc}
\end{figure*}

BSSs are identified as stars positioned to the left and above the main-sequence turnoff (MSTO) in the optical color-magnitude diagram (CMD). They were first discovered by \citet{Sandage1953} in the GC M3. The position of BSSs in the CMD suggests that they are more massive than the current cluster members. This is further supported by previous mass measurements \citep{Shara1997, Gilliland1998, Fiorentino2014}. However, GCs are completely devoid of gas, making any recent star formation practically implausible. As a result, the origin of BSSs is associated with a mechanism that increases the initial stellar mass through a rejuvenation process. The two primary formation mechanisms of BSSs are mass transfer (MT) from an evolved donor to a lower-mass binary companion \citep{McCrea1964} and stellar collisions leading to mergers in high-density environments \citep{Hills1976, Leonard1989}. The absence of a single formation scenario that explains all observed BSSs across different clusters necessitates identifying distinguishing characteristics for each formation scenario.

Chemical abundance is one such indicator that distinguishes the BSS formation scenario. Smooth particle hydrodynamics calculations reveal that stellar collisions produce merged objects with surface abundances matching the envelope of the most massive star \citep{Lombardi1995}. In contrast, binary mass transfer exposes and accretes the donor star's deeper layers, resulting in surface abundances indicative of partial hydrogen burning. For instance, in 47 Tuc, \citet{Ferraro2006} found many BSSs depleted in carbon (C) and oxygen (O), suggesting they accreted CNO-processed material from binary companions. Such chemical anomalies may indicate mass-transfer mechanisms. Notably, 47 Tuc shows a high prevalence of depleted BSSs, unlike clusters like M4, M30, and $\omega$ Centauri, where only a few stars show a C-O signature \citep{Lovisi2010, Lovisi2013, Mucciarelli2014}. Small numbers of CO-depleted BSSs suggest that mass transfer is either inefficient in GCs or C and O depletion is transient \citep{Ferraro2006}. 

\citet{Billi2023} employed high-resolution spectra of BSSs in NGC 3201 obtained with the Magellan Telescope to determine their rotational velocities. They determined the effective temperature and surface gravity using isochrone fitting on the CMD locations of the BSSs. In this work, we aim to determine the radial velocity, atmospheric parameters and chemical abundances of the BSSs using the synthetic spectral fitting method.

NGC 3201 is a low galactic latitude halo GC in the southern constellation of Vela. The right ascension (RA) and declination (DEC) of the cluster center are $10^h 17^m 36.82^s$ and $-46^{\circ}$ 24' 44.9", respectively. The cluster has an age $\approx$ $12.2 \pm 0.5$ Gyr \citep{Monty2018}, a distance $\approx$ 4.9 kpc, metallicity [Fe/H] $\approx$ $-1.59$ dex, and reddening $\approx$ 0.24 mag \citep{Harris1996}.

The article is structured as follows: \S~\ref{data} describes the data sets used in this study. The methodology is presented in \S~\ref{method}. The results of this study are described in \S~\ref{results&discussion}, while \S~\ref{conclusions} presents the conclusions.

\section{Data} \label{data}

In this work, we have used BSS spectra, along with their corresponding RA \& DEC and \( v \sin(i) \) values, which are openly available at Cosmic-lab \footnote{\url{http://www.cosmic-lab.eu/Cosmic-Lab/BSS_rotation.html}}. The spectra were acquired from the multi-object fiber system Michigan/Magellan Fiber System (M2FS), which feeds the double spectrograph MSPec mounted on the Magellan Clay Telescope at the Las Campanas Observatory in Chile. The spectra were acquired as part of the proposal CN2019A-15 with Principal Investigator Lorenzo Monaco. The M2FS instrument enables the simultaneous observation of up to 128 objects per spectrograph over a FOV of about 30$^{\prime}$ in diameter. The data were collected in February-March 2019 through 16 repeated exposures in the spectral region 5127-5184 \AA, with a spectral resolution of 18000. This resolution allowed us to adequately sample the first two lines of the Mg triplet at 5167.3 \AA\ and 5172.6 \AA. Two distinct fiber configurations were utilized to obtain spectra for roughly 200 targets in the direction of the cluster. Our analysis targeted objects with spectra that reached an optimal Signal-to-Noise ratio (S/N) of $\sim$ 40–50. A total of six exposures, each lasting 30 minutes, were acquired on the first night, while 10 exposures, each lasting 20 minutes, were obtained on the second night \citep{Billi2023}. \citet{Billi2023} provides a detailed description of the data reduction process and their methodology for determining radial velocity, cluster membership, atmospheric parameters, and \( v \sin(i) \) values for the BSSs.

\begin{figure}
\centering 
    \includegraphics[scale = 1]{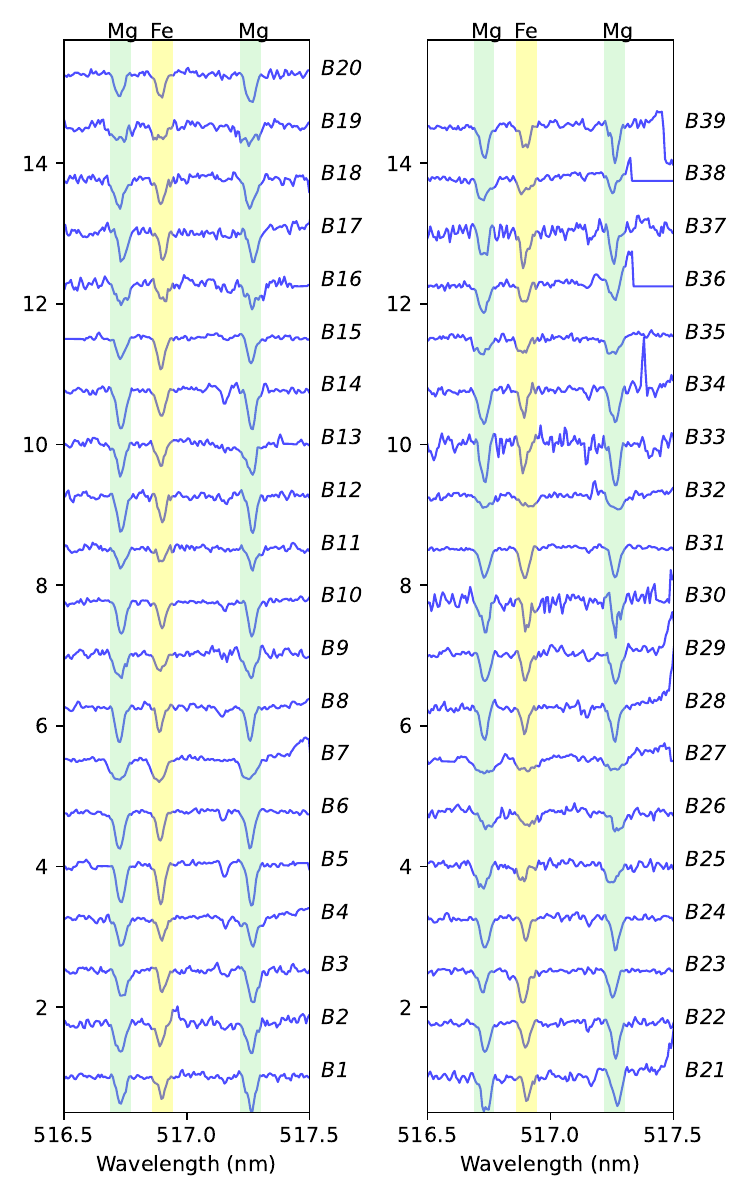}
    \caption{The normalized and radial velocity-corrected spectra of 39 BSSs (B1-B39) of the cluster NGC 3201. The Mg and Fe lines are shown with vertical bands.}
    \label{fig:ngc3201_bss_spectra}
\end{figure}

Out of the 67 BSSs analyzed by \citet{Billi2023}, we used 39 BSSs for which we were able to sample the Mg and Fe lines (see Table \ref{tab:ngc3201_analysis_lines_details}) for proper atmospheric parameters and chemical abundance estimation. The ID, RA, DEC, visual magnitude, and cluster membership probability for these 39 BSSs can be found in Table \ref{tab:bss_crossmatch_results}. The BSS IDs are named such that smaller numbers correspond to BSSs closer to the cluster's core. The Cosmic-Lab ID, RA and DEC are directly taken from the Cosmic-Lab website. The apparent visual magnitude (\lowercase{m}$_V$) is obtained by crossmatching the BSS\lowercase{s} positions with the photometric catalogue of the Stetson database \citep{Stetson2019}. The cluster membership probabilities (PROB) are obtained from \citet{Nadiello2018} and \citet{Vasiliev2021}. 

The spatial locations of these BSSs in different cluster regions can be seen in Figure \ref{fig:ngc3201_bss_loc}. Out of the total 39 BSSs used in this study, 13 are within the core radius ($R_c = 1.3'$), 13 lie between the core and half-light radii, and 13 are beyond the half-light radius ($R_h = 3.1'$) of NGC~3201 \citep{Harris1996}. The continuum-normalized and radial velocity-corrected spectra of the BSSs are given in Figure \ref{fig:ngc3201_bss_spectra}.

\begin{figure}
    \centering
    \includegraphics[scale=1]{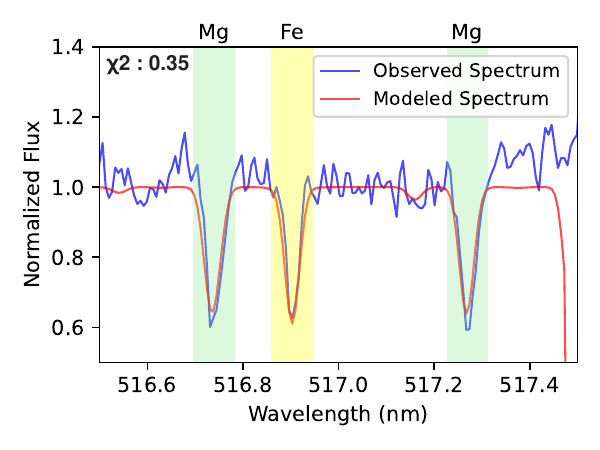}
    \caption{The observed spectrum, focusing on Mg and Fe lines, is fitted with a model. The observed spectrum is shown in blue, while the fitted model is shown in red. The light green and yellow regions represent the Mg and Fe lines.}
    \label{fig:modelling_example}
\end{figure}

\section{Methodology} \label{method}

We utilized \texttt{iSpec} \citep{Blanco2014, Blanco2019} for spectral analysis in our work. \texttt{iSpec} is an integrated spectroscopic software framework\footnote{\url{https://www.blancocuaresma.com/s/iSpec}} designed for determining astrophysical parameters and individual chemical abundances. In \texttt{iSpec}, we used the synthetic spectral fitting method to extract atmospheric parameters and chemical abundances from BSS spectra. This method compares the observed spectrum with synthetic spectra generated on-the-fly \citep{Valenti1996}, focusing on specific spectral features. Utilizing a least-squares algorithm, the differences between the synthetic and observed spectra are minimized to obtain optimal parameters.

The continuum points within a spectrum are identified by applying a median and maximum filter with varying window sizes. The median filter smooths out noise, while the maximum filter disregards deeper fluxes associated with absorption lines \citep{Blanco2014}. Subsequently, a polynomial fit is employed to model the continuum. Finally, the spectrum is normalized by dividing all flux values by the modelled continuum.

Initially, We determined the radial velocity using two lines of the Mg triplet at 5167.3 \AA and 5172.6 \AA. After calculating the radial velocities, the spectra were shifted to their rest wavelengths to ensure accurate alignment for further analysis.

{The synthetic spectrum was fitted to the observed spectra to determine atmospheric parameters and chemical abundances using \texttt{iSpec} \citep{Sarmento2020, Casamiquela2022}. The synthetic spectra were generated using the \texttt{ATLAS9.KuruczODFNEW} \citep{Castelli2003} model atmosphere with the \texttt{SYNTHE} \citep{Kurucz1993, Sbordone2004} radiative transfer code in \texttt{iSpec}, and the Gaia-ESO Survey line list \citep{Heiter2015, Blanco2019}. Solar abundances were adopted from \citep{Grevesse1998}. Figure \ref{fig:modelling_example} illustrates an example of model fitting where an observed spectrum of a BSS (B17) is compared against synthetic spectra, focussing on Mg and Fe lines. The observed spectrum (blue) is well-matched by the synthetic spectrum (red), achieving a chi-squared value of \( \chi^2 = 0.35 \). In general, the $\chi^2$ values for the spectral fitting of other BSSs were found to be in the range of 0-1.

To determine the atmospheric parameters, we performed synthetic spectral fitting in two steps, focusing on Mg lines in the first step and Fe lines in the second (see Table \ref{tab:ngc3201_analysis_lines_details}). The initial parameters for effective temperature (T$_{\text{eff}}$) and surface gravity ($\log g$) were set to 7700 K and 4 dex, respectively, based on the ranges quoted in previous work (6700–8700 K for T$_{\text{eff}}$ and 3.6–4.3 dex for $\log g$; \citep{Billi2023}). The metallicity ([M/H]) was fixed at $-1.59$, consistent with the cluster's [Fe/H] value \citep{Harris1996}. The  $v \sin i$ values were sourced from Cosmic-Lab. Initial values for microturbulence ($v_{\text{mic}}$) and macroturbulence ($v_{\text{mac}}$) were estimated using the built-in functions \texttt{estimate\_vmic} and \texttt{estimate\_vmac} in \texttt{iSpec}, which use empirical relations based on T$_{\text{eff}}$, $\log g$, and [M/H]. These initial values were 2.27 \kms\ for $v_{\text{mic}}$ and 26.73 \kms\ for $v_{\text{mac}}$.

Atmospheric parameters (T$_{\text{eff}}$ and $\log g$) were derived independently from Mg and Fe lines. The final T$_{\text{eff}}$ and $\log g$ values listed in Table \ref{tab:atmos_params_results} for each BSS were computed as the means of the values derived from Mg and Fe lines, with the standard deviations representing the errors.

Using the final atmospheric parameters, we re-ran the spectral fitting to derive the values of [Mg/Fe] and [Fe/H]. The uncertainties in [Mg/Fe] and [Fe/H] were calculated by propagating the errors in T$_{\text{eff}}$ and $\log g$. For this, the abundances were re-computed by varying each atmospheric parameter (T$_{\text{eff}}$, $\log g$) within its upper and lower uncertainty limits while keeping the others fixed. The resulting variations in the abundances were combined in quadrature to estimate the final uncertainties. Given the limited number of spectral lines available, these results should be approximate estimates.

\begin{table}
    \centering
    \caption{M\lowercase{g} and F\lowercase{e} spectral lines used in this study; \lowercase{log $gf$} Denotes the Logarithm of the Product of Statistical Weight and Oscillator Strength}
    
    \label{tab:ngc3201_analysis_lines_details}
    \begin{tabular}{ccc}
    \toprule
    Element & Peak Wavelength (nm) & $\log gf$ \\
    \midrule 
    Mg & 516.732 & -0.931 \\
                        & 517.268 & -0.450 \\
    \midrule
    Fe                 & 516.903 & -1.000 \\
    \bottomrule 
\end{tabular}

\end{table}

\section{Results and discussion}\label{results&discussion}

Using the methodology provided in \S~\ref{method}, we performed spectral analysis of the BSSs in NGC 3201. The radial velocities, atmospheric parameters (T$_{\text{eff}}$ and log $g$) and chemical abundances derived from the spectra for all 39 BSSs are provided in Table \ref{tab:atmos_params_results}. 

\begin{table*}
\centering 
\begin{center}
\captionof{table}{Atmospheric parameters and chemical abundances of the BSS\lowercase{s}. The values of \lowercase{$vsin(i)$} is taken from Cosmic-lab.}
\label{tab:atmos_params_results}
\begin{tabular}{|ccccccc|}
\hline
BSS ID & RV (\kms)   & $v$ sin($i$) (\kms) & T${_\text{eff}}$ (K) & log $g$  & [Fe/H] & [Mg/Fe]  \\ \hline 

B1	&	503.52	&	5	$\pm$	2	&	7112	$\pm$	529	&	4.5	$\pm$	0.7	&	-1.47	$\pm$	0.22	&	-0.14	$\pm$	0.62	\\
B2	&	502.21	&	12	$\pm$	2	&	7830	$\pm$	487	&	4.9	$\pm$	0.1	&	-1.22	$\pm$	0.11	&	1.39	$\pm$	0.57	\\
B3	&	482.51	&	13	$\pm$	2	&	6951	$\pm$	726	&	4.9	$\pm$	0.1	&	-1.43	$\pm$	0.33	&	0.49	$\pm$	0.83	\\
B4	&	492.60	&	9	$\pm$	1	&	6910	$\pm$	517	&	3.9	$\pm$	0.9	&	-1.32	$\pm$	0.61	&	0.32	$\pm$	0.72	\\
B5	&	493.97	&	5	$\pm$	1	&	6260	$\pm$	629	&	2.6	$\pm$	0.9	&	-1.28	$\pm$	0.34	&	-0.07	$\pm$	0.56	\\
B6	&	498.02	&	5	$\pm$	1	&	6962	$\pm$	118	&	3.6	$\pm$	0.9	&	-1.04	$\pm$	0.56	&	1.21	$\pm$	0.46	\\
B7	&	499.53	&	25	$\pm$	2	&	6780	$\pm$	617	&	3.2	$\pm$	0.9	&	-1.21	$\pm$	0.42	&	0.05	$\pm$	0.44	\\
B8	&	507.22	&	5	$\pm$	1	&	7772	$\pm$	771	&	4.5	$\pm$	0.7	&	-1.13	$\pm$	0.28	&	1.36	$\pm$	0.81	\\
B9	&	493.97	&	16	$\pm$	3	&	7724	$\pm$	146	&	4.5	$\pm$	0.7	&	-1.40	$\pm$	0.23	&	1.06	$\pm$	0.23	\\
B10	&	496.45	&	5	$\pm$	1	&	6545	$\pm$	634	&	3.4	$\pm$	0.9	&	-1.63	$\pm$	0.46	&	-0.43	$\pm$	0.65	\\
B11	&	497.79	&	16	$\pm$	3	&	8283	$\pm$	371	&	4.8	$\pm$	0.3	&	-1.76	$\pm$	0.11	&	0.52	$\pm$	0.39	\\
B12	&	493.97	&	5	$\pm$	2	&	6659	$\pm$	697	&	3.9	$\pm$	0.9	&	-1.35	$\pm$	0.53	&	0.46	$\pm$	1.02	\\
B13	&	509.48	&	9	$\pm$	3	&	7040	$\pm$	146	&	4.9	$\pm$	0.1	&	-1.15	$\pm$	0.08	&	0.36	$\pm$	0.16	\\
B14	&	497.79	&	5	$\pm$	1	&	6555	$\pm$	277	&	3.4	$\pm$	0.6	&	-1.65	$\pm$	0.31	&	0.40	$\pm$	0.24	\\
B15	&	498.02	&	14	$\pm$	2	&	6714	$\pm$	723	&	2.9	$\pm$	0.9	&	-1.50	$\pm$	0.51	&	-1.32	$\pm$	0.75	\\
B16	&	488.24	&	20	$\pm$	2	&	7388	$\pm$	772	&	4.9	$\pm$	0.1	&	-1.38	$\pm$	0.16	&	0.05	$\pm$	1.12	\\
B17	&	490.68	&	5	$\pm$	2	&	6717	$\pm$	856	&	3.3	$\pm$	0.9	&	-1.96	$\pm$	0.52	&	-0.80	$\pm$	0.72	\\
B18	&	498.02	&	15	$\pm$	2	&	6899	$\pm$	741	&	3.4	$\pm$	0.8	&	-1.77	$\pm$	0.59	&	-0.04	$\pm$	0.52	\\
B19	&	499.70	&	36	$\pm$	4	&	7735	$\pm$	825	&	5.0	$\pm$	0.0	&	-1.40	$\pm$	0.12	&	-0.03	$\pm$	0.98	\\
B20	&	501.61	&	10	$\pm$	3	&	7202	$\pm$	701	&	3.7	$\pm$	0.9	&	-1.22	$\pm$	0.66	&	-0.20	$\pm$	0.77	\\
B21	&	495.69	&	5	$\pm$	2	&	6490	$\pm$	409	&	4.0	$\pm$	0.8	&	-1.67	$\pm$	0.39	&	0.65	$\pm$	0.66	\\
B22	&	495.88	&	5	$\pm$	2	&	6953	$\pm$	712	&	4.0	$\pm$	0.9	&	-1.47	$\pm$	0.50	&	-0.08	$\pm$	0.97	\\
B23	&	503.38	&	9	$\pm$	1	&	7010	$\pm$	591	&	3.9	$\pm$	0.9	&	-1.24	$\pm$	0.52	&	-0.69	$\pm$	0.59	\\
B24	&	490.15	&	6	$\pm$	2	&	6864	$\pm$	307	&	4.3	$\pm$	0.9	&	-1.60	$\pm$	0.43	&	0.37	$\pm$	0.58	\\
B25	&	498.02	&	17	$\pm$	3	&	8690	$\pm$	673	&	4.9	$\pm$	0.2	&	-1.23	$\pm$	0.12	&	1.89	$\pm$	1.06	\\
B26	&	493.97	&	25	$\pm$	3	&	8373	$\pm$	551	&	3.0	$\pm$	0.7	&	-2.36	$\pm$	0.49	&	0.48	$\pm$	0.52	\\
B27	&	495.69	&	35	$\pm$	4	&	7420	$\pm$	825	&	4.4	$\pm$	0.5	&	-1.27	$\pm$	0.25	&	0.39	$\pm$	0.91	\\
B28	&	504.13	&	5	$\pm$	2	&	7000	$\pm$	343	&	3.9	$\pm$	0.9	&	-1.27	$\pm$	0.58	&	0.35	$\pm$	0.50	\\
B29	&	492.60	&	5	$\pm$	2	&	6823	$\pm$	788	&	3.9	$\pm$	0.9	&	-1.14	$\pm$	0.57	&	-0.02	$\pm$	1.08	\\
B30	&	503.52	&	5	$\pm$	2	&	6898	$\pm$	890	&	3.6	$\pm$	0.9	&	-1.13	$\pm$	0.51	&	-0.03	$\pm$	1.42	\\
B31	&	498.02	&	10	$\pm$	1	&	6564	$\pm$	900	&	3.0	$\pm$	0.8	&	-1.38	$\pm$	0.51	&	-0.70	$\pm$	0.92	\\
B32	&	498.36	&	26	$\pm$	2	&	8419	$\pm$	442	&	4.0	$\pm$	0.7	&	-1.60	$\pm$	0.60	&	0.69	$\pm$	0.63	\\
B33	&	501.61	&	5	$\pm$	2	&	7365	$\pm$	382	&	4.9	$\pm$	0.2	&	-1.07	$\pm$	0.09	&	0.80	$\pm$	0.38	\\
B34	&	501.84	&	5	$\pm$	1	&	7660	$\pm$	666	&	4.3	$\pm$	0.8	&	-1.28	$\pm$	0.28	&	1.37	$\pm$	0.71	\\
B35	&	499.53	&	27	$\pm$	3	&	7877	$\pm$	186	&	4.2	$\pm$	0.8	&	-1.69	$\pm$	0.56	&	1.03	$\pm$	0.52	\\
B36	&	495.69	&	13	$\pm$	2	&	7964	$\pm$	622	&	4.4	$\pm$	0.9	&	-1.43	$\pm$	0.32	&	1.61	$\pm$	0.78	\\
B37	&	507.22	&	5	$\pm$	2	&	7217	$\pm$	400	&	3.8	$\pm$	0.9	&	-1.69	$\pm$	0.67	&	-0.79	$\pm$	0.46	\\
B38	&	507.22	&	17	$\pm$	3	&	8415	$\pm$	367	&	4.6	$\pm$	0.6	&	-1.59	$\pm$	0.26	&	1.61	$\pm$	0.53	\\
B39	&	501.61	&	5	$\pm$	2	&	7059	$\pm$	457	&	4.2	$\pm$	0.9	&	-0.97	$\pm$	0.79	&	0.49	$\pm$	0.71	\\
\hline 
\end{tabular}
\end{center}
\end{table*}

Figure \ref{fig:rv_hist} shows a histogram of radial velocities obtained for the BSSs. We performed a Gaussian fit to determine the average radial velocity for the BSSs as 498.0 $\pm$ 5.3 \kms. Comparing this with the cluster's radial velocity of 494.0 $\pm$ 0.2 \kms \citep{Harris1996}, the BSSs' average radial velocity is similar within error.

\begin{figure}
    \centering
    \includegraphics[scale=1]{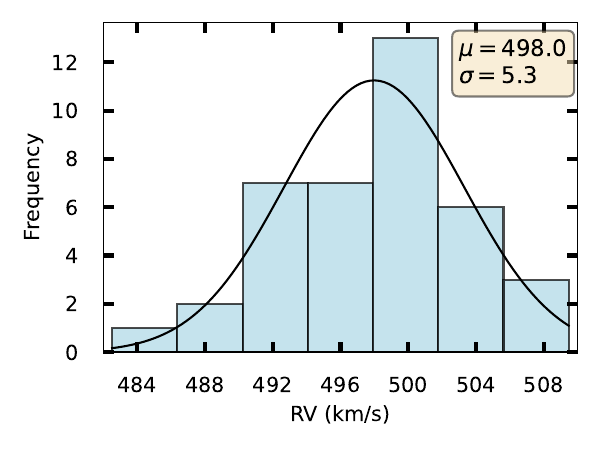}
    \caption{Histograms of Radial velocity estimated for the 39 BSSs. }
    \label{fig:rv_hist}
\end{figure}

\begin{figure}
    \centering
    \includegraphics[scale=0.8]{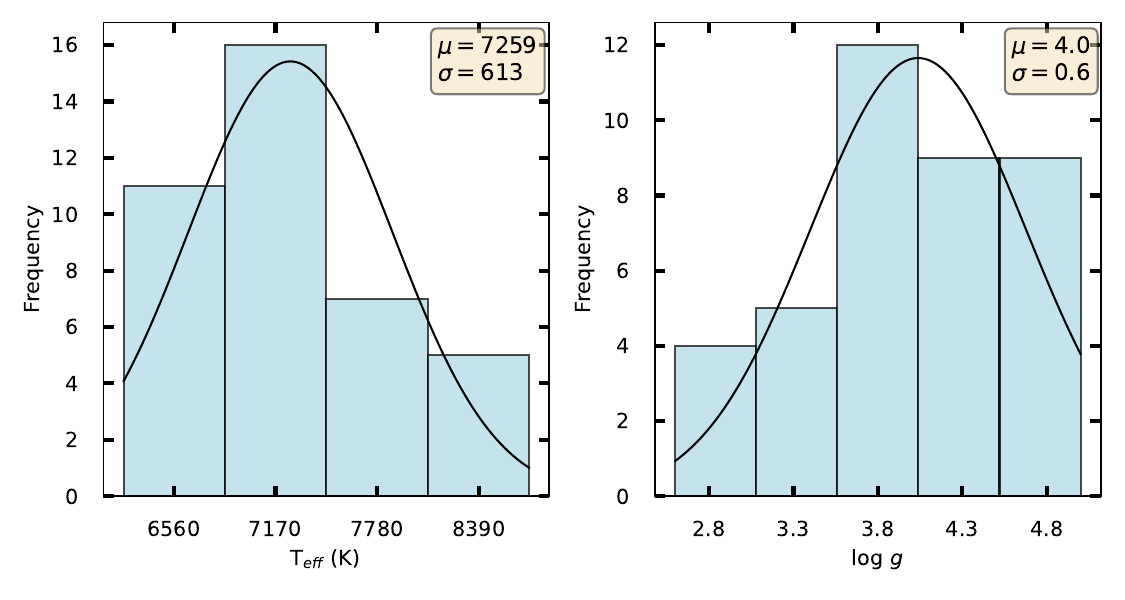}

    \caption{Histograms of T$_{\text{eff}}$ and log $g$ estimated for the BSSs.}
    \label{fig:atmosparams_hist}
\end{figure}

\begin{figure}
    \centering
    \includegraphics[scale=0.8]{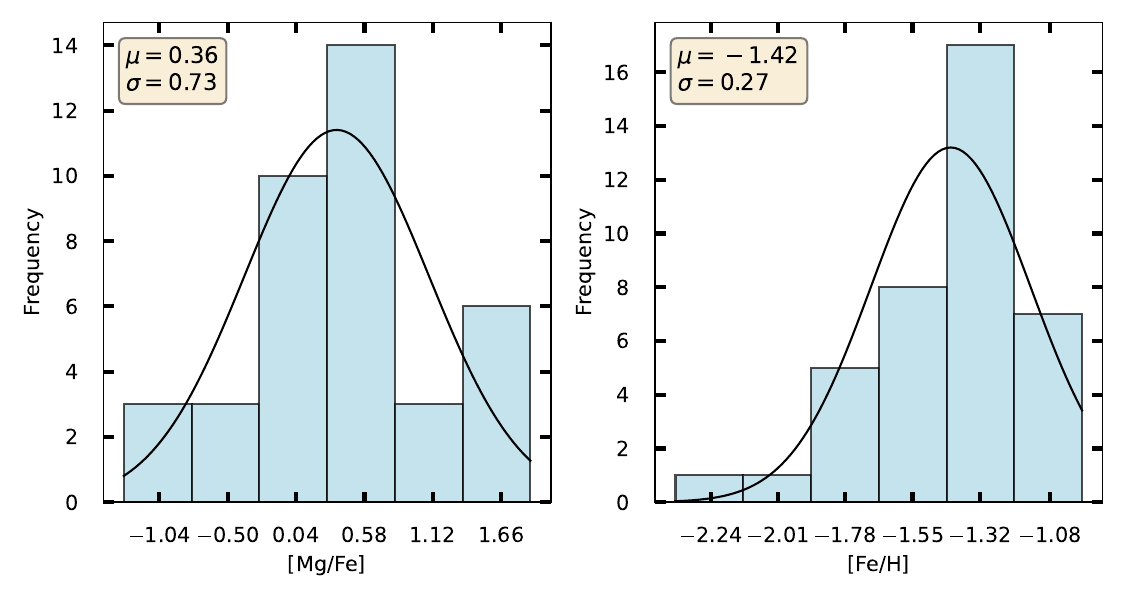}
    \caption{Histograms of [Mg/Fe] and [Fe/H] values estimated for BSSs.}
    \label{fig:abundance_hist}
\end{figure}

Figure \ref{fig:atmosparams_hist} provides insights into the obtained T$_{\text{eff}}$ and log $g$ values. The range of  T$_{\text{eff}}$ and log g for the BSS is 6260 - 8690 K and 2.6 - 5.0, respectively.  \citet{Billi2023} quoted the T$_{\text{eff}}$ and log $g$ in the range of 6700-8700 K and 3.6-4.3, respectively, for the BSSs in NGC 3201. Our derived values are also nearly in the same range. The Gaussian fitting to the histogram provides a mean T$_{\text{eff}}$ and log $g$ of the BSSs as 7259 $\pm$ 613 K and 4.0 $\pm$ 0.6 dex, respectively.

Figure \ref{fig:abundance_hist} provides histogram plots of the obtained chemical abundances of [Mg/Fe] and [Fe/H]. We performed Gaussian fitting on these distributions to determine the average values of [Mg/Fe] and [Fe/H]. The mean values obtained are [Mg/Fe] = 0.36 $\pm$ 0.73 and [Fe/H] = -1.42 $\pm$ 0.27. \citet{Munoz2013} performed a detailed chemical abundance analysis of eight red giants in NGC 3201 using high-resolution spectroscopy. They found mean values of [Fe/H] = -1.53 $\pm$ 0.01 and [Mg/Fe] = 0.38 $\pm$ 0.03. These values are comparable with errors in our measurements for the BSSs. However, we observe significant dispersion in our abundance values, likely due to the lower SNR, resolution, and the limited number of available lines. The obtained mean [Fe/H] value of -1.42 $\pm$ 0.27 is consistent with the cluster’s [Fe/H] value of -1.59 within error. 

As mentioned in \S~\ref{introduction}, two main scenarios have been proposed for forming BSS: mass transfer and stellar collisions leading to mergers. Distinguishing which of these processes gave rise to BSS is challenging. However, chemical abundance analysis of BSS may be the most effective method for differentiating these two formation scenarios. Collisional BSSs are not expected to exhibit abundance anomalies. In contrast, mass transfer BSSs typically show depleted surface abundances of $\alpha$-elements such as carbon (C), oxygen (O), and magnesium (Mg) \citep{Lovisi2013}. Our derived values for [Mg/Fe] have large uncertainty, which prevents us from making definitive statements about the formation mechanisms of these BSSs.

While we attempted to identify correlations within the current dataset, no statistically significant relationships could be established. We also looked into the color histogram for the BSSs but no relationship was observed. Nevertheless, the robustness of the present analysis suggests that future studies with broader spectral coverage in NGC 3201 may yield important insights into potential correlations between the chemical abundances and atmospheric parameters of BSSs.

\section{Conclusions} \label{conclusions}

 We analyzed the spectra of 39 BSSs in the globular cluster NGC 3201 and derived radial velocity, atmospheric parameters, and chemical abundance of [Mg/Fe] and [Fe/H]. The present study concludes the following.

\begin{enumerate}
    \item  The radial velocity of the 39 BSSs is determined using the Mg and Fe spectral lines, found in the 482.51 - 509.48 km/s range with a mean value of 498.0 $\pm$ 5.3 km/s.
    \item The T$_{eff}$  and log $g$ values are found in the range of 6260 - 8690 K and 2.6 - 5.0, respectively, for the BSSs. 
    \item The abundance values of Mg and Fe are derived using the same lines. The mean value of [Mg/Fe] is estimated as 0.36 $\pm$ 0.73, while the value of [Fe/H] is estimated as -1.42 $\pm$ 0.27. 
\end{enumerate}

\section*{ACKNOWLEDGEMENTS}
We are grateful to Cosmic-Lab, especially Alex Billi and FR. Ferraro, for providing data on BSSs in NGC 3201 and offering insights into spectral corrections.

\bibliography{rm-extenso}
\bibliographystyle{rmaa}

\end{document}